\documentclass[reprint,floatfix,nofootinbib]{revtex4-1}
\usepackage{color}

\newcommand{\rem}[1]{}

\usepackage{graphicx}
\usepackage{dcolumn}
\usepackage{bm}
\usepackage{amsmath,amssymb,amsthm} 
\usepackage{makeidx}
\usepackage{amsfonts}
\usepackage{comment}
\usepackage[ansinew]{inputenc}
\usepackage[usenames,dvipsnames]{pstricks}
\usepackage{subfigure}
\usepackage{epsfig}
\usepackage{braket}
\usepackage{float}

\begin{document}

\preprint{APS/123-QED}

\title{Integrability of the hyperbolic reduced Maxwell-Bloch equations for strongly correlated Bose-Einstein condensates}

\author{ Alexis Arnaudon}
\email{alexis.arnaudon@imperial.ac.uk}
\author{John D. Gibbon}
\affiliation{Department of Mathematics, Imperial College, London SW7 2AZ, UK.}

\date{\today}

\begin{abstract}
    We derive and study the hyperbolic reduced Maxwell-Bloch equations (HRMB), a simplified model for the dynamics of strongly correlated Bose-Einstein condensates (BECs), and in particular for the interaction between the BEC atoms and its evaporated atoms under the strong interactions.  
    This equation is one among four which are proven to be integrable via the existence of a Lax pair, and thus the method of inverse scattering transform.  
    Another equation is the reduced Maxwell-Bloch equation of quantum optics and the two others do not have physical applications yet. 
    By studying the linear stability of the constant solutions of these four equations we observe various regimes, from stable, to modulational unstable and unstable at all frequencies. 
    The finite dimensional reduction of the RMB equations is also used to give more insight into the constant solutions of these equations. 
    From this study, we find that the HRMB equation arising from strongly correlated BECS is stable under the particular condition that the transition rate of evaporation is not too large compared to the number of evaporated atoms.  
    We then derive explicit soliton solutions of the RMB equations and use numerical simulations to show collisions of solitons and kink solitons. 
\end{abstract}

\pacs{todo }


\maketitle

\section{Introduction}

The excitation and propagation of solitons in Bose-Einstein condensates (BECs) has been an active area of study for a number of years. 
Two reviews cover the more general area of BECs \cite{dalfovo1999theory,giorgini2008theory} while two more put greater emphasis on soliton excitation \cite{kivshar1998dark,frantzeskakis2010dark}. 
Experimental studies of strongly correlated BECs have very recently become possible \cite{burger1999dark,strecker2002formation,wuster2007quantum,xu2006observation} and new phenomena have emerged \cite{makotyn2014universal,kira2015coherent}. 
The fundamental parameter in these experiments is the correlation length between atoms of the BEC, represented as a scattering length $a_\mathrm{scatt}$.  
An abrupt change of this $a_\mathrm{scatt}$ from small to large can create a stable intermediate state that would usually have evaporated due to the strong interactions \cite{makotyn2014universal}. 
The excited atoms that are still interacting with the BEC form the normal component of the BEC. 

A mathematical description of strongly correlated BECs was recently initiated by Kira \cite{kira2014excitation, kira2015hyperbolic} who derived the so-called hyperbolic Bloch equations (HBE) to model the excited atoms of the BEC. 
The derivation of this equation follows the one for the semiconductor Bloch equations, based on a cluster expansion approach of the normal component of the strongly correlated BEC, or the electrons in the semiconductor. 
The application of this cluster expansion method is not directly applicable to the BEC dynamics because all orders in the cluster expansion would be required. 
This difficulty is circumvented by the application of a non-unitary transformation that uses the normal component alone by representing the BEC as the vacuum state. 
The expansion can then be carried out to arbitrary orders. 
In this work, we will only consider the first order terms which describe the singlet dynamics and show a simplified derivation. 
We refer the reader to \cite{kira2014excitation, kira2015hyperbolic,kira2015coherent} for a complete description of this method for the BECs, and to \cite{kira2011semiconductor} for its initial use for semiconductors in quantum optics. 

Although the physics is completely different BECs and semiconductors, the hyperbolic Bloch equation (HBE) and the semiconductor Bloch equation (SBE) share the same structure. 
The only difference lies in a minus sign which transforms the Bloch sphere into a hyperboloid on which the solutions evolve. 
A second important difference is in the coupling of these equations to the external dynamics of light, or BEC. 
In the case of semiconductor optics, the SBE is coupled electromagnetically through a wave equation for the electric field which contains a term dependent on the state of the semiconductor. 
Recall that the HBE does not describe the BEC dynamics which thus requires a coupling to the Gross-Pitaevskii equation (GP) to describe the full dynamics including the BEC. 
In fact, the BEC wave function replaces the wave equation of the electric field and the coupling is performed via a source term that describes the local loss or gain of atoms in the BEC. 
        
In this paper, we derive a particular approximation of the coupled HBE and GP equations that will be shown to be completely integrable. 
This approximation roughly corresponds to considering solutions with small amplitude with respect to the average amplitude of the BEC. 
To emphasise the parallel with optics, a similar approximation made more than 40 years ago yielded the reduced Maxwell-Bloch equations (RMB) in quantum optics, a completely integrable equation\,: see \cite{eilbeck1973solitons,gibbon1973soliton,caudrey1974exact,dodd1982solitons} and references therein.  
The resulting equations in this context of BECs will be called the hyperbolic reduced Maxwell-Bloch equations (HRMB). 

In fact, we will show that the RMB and the HRMB equations are two equations among a group of four inequivalent integrable equations members of the first negative flow of the AKNS hierarchy \cite{zakharov1972exact,ablowitz1973nonlinear}. 
All these RMB equations can be integrated via the inverse scattering transform and contain soliton solutions. 
The solutions are either pulses or kinks with possibly two directions of propagation, depending on the RMB equation. 
Another interesting solution is the constant solution which reduces the original PDE to the non-dissipative Lorenz 63 model \cite{lorenz1963deterministic}. 
The stability of this constant solution shows that the RMB equations of semiconductor optics admit a regime of modulational instability and that the HRMB equation is stable for all wavelength provided the number of atoms in the normal component is large enough compared to the interaction rate with the BEC. 

We wish to emphasise that the aim of this paper is not to present a complete, detailed derivation and description of the HRMB equations, but to rather make a new connection between two areas of science, namely strongly correlated Bose-Einstein condensates and integrable systems. 


\section{Physical Derivation}

\subsection{Review of the RMB equations}

The Maxwell-Bloch equations first appeared in quantum optics in the context of the phenomenon called self-induced transparency\,: see for example \cite{bullough1974general,maimistov1990present} for reviews on this topic.  
More recently in quantum semiconductor optics, a more general form of these equations that can be reduced to the Maxwell-Bloch equations after neglecting the extra higher order terms is used. 
We refer to \cite{kira2011semiconductor} for a recent monograph on this topic. 
For the purpose of this work, we will prefer the semiconductor description of the Maxwell-Bloch equations as their derivation uses the same method that for the derivation of the HBE equations in \cite{kira2015hyperbolic}.  
Let us first recall the SBE in its simplest form
\begin{align}
    \begin{split}
    i \dot p &= \omega_0 p + (2f-1)\Omega\\
    \dot f &= -2\, \mathrm{Im}\left (  \Omega p ^*\right )\, ,
    \end{split}
    \label{SBE}
\end{align}
where $p$ is a complex field representing the transition amplitude between the state of an electron and a state of a hole.
The scalar field $f$ is the occupation number of the electrons that varies between $-1$ and $1$. 
The complex number $\Omega$ is the Rabi energy, which is proportional to the electric field applied to the system. 
The equations \eqref{SBE} conserve the quantity 
\begin{align}
    \eta = \left (f-\frac12\right )^2+ |p|^2\, , 
\end{align}
which represents the Bloch sphere of radius $\sqrt{\eta}$. 
The equations \eqref{SBE} are already simplifications of the complete model because they incorporate the sharp line approximation. 
This amounts to writing the equations with a single resonance frequency $\omega_0$ and no frequency averaging with a response function.  
These equations can be derived with the cluster expansion approach, a method similar to the BBGKY hierarchy that allows the computation of the many-body interactions between electrons up to some order. 
Equations \eqref{SBE} only contain singlet terms and the more physically realistic doublet or triplet dynamics have been neglected.
Similar interesting phenomena occur in semiconductor quantum optics where the Bloch equations \eqref{SBE} are coupled to the standard Maxwell wave equation. 
The result is the semiconductor Maxwell-Bloch equations, where the wave equation for the electric field $E$ is coupled to $p$ via a small material parameter $\alpha_0$. 
The smallness of $\alpha_0$ together with the use of short intense pulses allows one to neglect the backscattering of waves in the Maxwell equation. 
The resulting wave equation is 
\begin{align}
    E_{t} + c E_{x} = \alpha_{0}p\,,
    \label{maxwell-RMB}
\end{align}
where $c$ the speed of light. 
Having removed the left-travelling waves, the resulting set of equations are called the reduced Maxwell-Bloch equation (RMB). 
In quantum optics, these equations govern the electric field, transition amplitude and occupation number variables before the slowly varying envelope approximation of McCall and Hahn \cite{mccall1969} are applied to produce the self-induced transparency (SIT) equation \cite{lamb1971}. 
The integrability of the RMB system and its generalisation to the hyperbolic case lies at the heart of this paper. 
We refer the reader to \cite{bullough1974general, eilbeck1973solitons,gibbon1973soliton,caudrey1974exact,dodd1982solitons} and references therein for more details on the derivation of these equations. 


\subsection{The hyperbolic Bloch equations}

The hyperbolic counterpart of the Bloch equations can be derived in the context of a strongly interacting BEC and is called the hyperbolic Bloch equation (HBE). 
We will expose below a simplified derivation of the HBE that can be found in substantial details in \cite{kira2015hyperbolic}. 
This equation appears in the context of strongly correlated Bose-Einstein condensates, where the internal correlations between the atoms in the BEC are strong enough to eject enough atoms to trigger the evaporation of the BEC. 
In particular cases (see \cite{kira2015coherent}), it is possible to obtain a state where both the BEC and the ejected atoms, or so-called normal component of the BEC, persists and interact non-linearly with the BEC itself. 
This strongly interacting regime is characterised by the limit $a_\mathrm{scatt}\to \infty$ which in practice means that the scattering length is saturated.  
This scattering length is experimentally controlled by the application of an external uniform magnetic field that triggers the so-called Feshbach resonance\,: see \cite{kira2015hyperbolic} or \cite{dalfovo1999theory,giorgini2008theory} for more details on this interaction potential. 

The HBE equation aims at describing the dynamics of atoms ejected from the BEC but remaining in interaction with it. 
Kira \cite{kira2015hyperbolic} used the method of cluster-expansion developed for semiconductor physics (see the monograph \cite{kira2011semiconductor}) to describe the dynamics of these atoms. 
This method can be used only for the ejected atoms in the strongly interacting regime because the expansion for the BEC itself must contain all orders in particle interactions.
In order to take only into account the normal component, a non-unitary transformation is applied to the BEC wave function to replace the BEC with a ground state and to concentrate only on the dynamics of the atoms in the normal component. 
This technique developed in \cite{kira2015coherent} is called the excitation picture and allows the precise study of the normal component of the strongly correlated BEC. 

In more details, the derivation begins with the quantum mechanical description of a Bose gas using the bosonic operators $B_k$ with wavenumber $k$.
The commutation relations are
\begin{align}
    [B_k,B^\dagger_{k'} ] = \delta_{k,k'}, \quad     
    [B_k,B_{k'} ] = [B_k^\dagger,B^\dagger_{k'} ]= 0\, , 
    \label{B-comm}
\end{align}
where $B_k^\dagger$ stands for the conjugate transpose of $B_k$ and the bracket is the commutation operator, which would be the anti-commutation for the fermionic operators.
Each of these operators has a momentum $\hbar k$ and an energy 
\begin{align}
    E_k = \frac{\hbar^2 k^2}{2m}\, , 
\end{align}
where $m$ is the mass of the Bosons and $\hbar$ the Planck constant. 
The case $k=0$ is the ground state of the system and will correspond to the condensed atoms in the BEC. 
All the states with $k\neq 0$ form the evaporated atoms. 

For the present discussion, we are instead interested by the dynamics of the expected value of these operators which will correspond to observable quantities.
The dynamical equation being nonlinear, computing an exact equation is impossible but we can approximate it with the cluster expansion approach.
First, the expectation of a product of operators can be written 
\begin{align}
    \Braket{B_kB_{k'}^\dagger}= \Braket{B_k}\Braket{B_{k'}^\dagger} + \Delta \Braket{B_kB_{k'}^\dagger}\,  , 
    \label{expansion}
\end{align}
where the last term is the statistical correlation between the two operators. 
The non-unitary transformation simplifies these decompositions to yield the two fields
\begin{align}
    f_k = \Braket{B_k^\dagger B_k} &= \Delta \Braket{B_k^\dagger B_k} \label{f-def}\\
    p_k = \Braket{B_k B_{-k}} &= \Delta \Braket{B_k B_{-k}}\label{p-def}\, . 
\end{align}
The real-valued quantity $f_k$ represents the number of atoms in the state $k$ and the complex quantity $p_k$ describes the transition of pairs of atoms ejected from the BEC state $B_0$ to two evaporated state with opposite momenta. 
The complex conjugate of this quantity describes the reverse process. 
For a more accurate description of the dynamic of evaporated atoms, higher order processes should be considered in the cluster expansion. 

We then derive the dynamical equation of these observables $f_k$ and $p_k$, using two approximations. 
\begin{itemize}
    \item We neglect all correlations higher than the singlets and doublets, and
    \item we approximate the inter-atomic interactions by a contact potential\,; that is, a Dirac delta function, or in Fourier space 
        \begin{align}
            V_k = \frac{4\pi \hbar^{2}}{ m} a_\mathrm{scatt}\, . 
        \end{align}
\end{itemize}
The second approximation allows us to consider a single wavenumber, so we can use only the following four Bosonic operators $B_+,B_-,B^\dagger_+$ and $B^\dagger_-$, and we can drop the $k$ subscripts for $f$ and $p$. 
With these approximations, the Hamiltonian of this system used in \cite{kira2015hyperbolic,kira2015coherent} is, in the excitation picture, given by 
\begin{align}
    \begin{split}
    \hat H_\mathrm{ex} &= \omega_0 B^\dagger_+ B_+ + \Omega ( B_- B_+ + B^\dagger_- B^\dagger_+ ) \, , 
    \end{split}
    \label{Hex}
\end{align}
where $\omega_0$ is the transition energy and 
\begin{align}
    \Omega= VN_c= 4\pi \hbar^2m^{-1} a_\mathrm{scatt}N_c
\end{align}
is the quantum-depletion source, proportional to the number of condensed atoms 
\begin{align}
    N_c= N_\mathrm{tot}- f\, .
    \label{Nc}
\end{align}
Here, $N_\mathrm{tot}$ is the total number of atoms in the system, taken to be constant.
We did not write the normal constant part of the Hamiltonian and only remark that the first term is the usual quantum harmonic oscillator, and the second term is a source and depletion term for the pair of atoms with opposite momenta. 

From this Hamiltonian, we compute the dynamic of an observable $O$ using the Liouville-van Neumann equation
\begin{align}
    i\hbar \frac{d}{dt} \braket{\hat O} = \Braket{\left [\hat H_\mathrm{ex},\hat O\right]}\, , 
\end{align}
and the fact that $f_-= f_+$ in \eqref{f-def} to obtain the simplified HBE equation 
\begin{align}
    \begin{split}
    i \dot p &= \omega_0 p + (2f+1) \Omega \\
    \dot f &= 2\,\mathrm{Im}\left (\Omega p^*\right )\, .
    \end{split}
    \label{HBE}
\end{align}
This equation corresponds to a simplification of the complete equations (101)-(102) of \cite{kira2015hyperbolic}.
If one uses anti-commutators in \eqref{B-comm}, one can check that we recover the semiconductor Bloch equation \eqref{SBE}.
Due to this sign flip in the $f$ equation, the hyperboloid 
\begin{align*}
    \eta = \left (f+\frac12\right )^2- |p|^2
\end{align*}
is preserved by the solution instead of the sphere for the SBE equations \eqref{SBE}.  


\subsection{Coupling with the Gross-Pitaevskii equation}

The next step in the derivation of the HRMB equations is to couple the HBE equations \eqref{HBE} with the Gross-Pitaevskii (GP) equation to include the internal BEC dynamics. 
Recall that in the case of the SBE \eqref{SBE}, the coupling with the Maxwell equation is achieved using the Rabi frequency and the electric field. 
In the case of BEC, the number of condensed atom plays the role of the electric field and the GP equation of the wave equation. 

The first approximation is the standard local-density approximation (LDA) which consists of studying the BEC dynamics locally, thus neglecting the exterior trapping potential and use the approximation of locally homogeneous BEC.  
The second approximation used here is to consider a one-dimensional condensate, that could still be valid in appropriate experiments\,: see for example \cite{burger1999dark}. 
The coupling between the BEC dynamics and the HBE is implemented as a source term in the GP equation, which reads
\begin{align}
    i\hbar \psi_t + \alpha \psi_{xx}+\beta \psi|\psi|^2  = i\beta \mathrm{Im}( p^*) \psi\, ,
    \label{GP}
\end{align}
where $\psi$ is a complex-valued wavefunction, $\alpha= \hbar^{2}/2m$, $\beta =8\pi a_\mathrm{scatt} \alpha $ and $m$ the mass of a boson. 
First, recall that the interaction length $a_\mathrm{scatt}\to \infty$ for strongly interacting BEC. 
Let us now write the GP equation in amplitude phase variables using the Madelung transformation $\psi(x,t) = \sqrt{n(x,t)}\exp\{i\phi(x,t)\}$ for the amplitude $n(x,t)$ and the phase $\phi(x,t)$
\begin{subequations}
\begin{align}
    n_t &+ 2(n\phi_x)_x=\beta \mathrm{Im}(p^*)n \label{n-eq} \\
    \phi_t &= \alpha\left ( \frac{ (\sqrt n)_{xx}}{\sqrt{n}} - \phi_x^2\right )  +2\beta n\label{phi-eq}\, .
\end{align}
\end{subequations}
Using the LDA, we can decompose the amplitude such that there is a constant background $n_0$ with a small perturbation $n_1(x,t)$, i.e.  $n(x,t)=n_0 + n_1(x,t)$, with $|n_1|\ll n_0$.
The steady solution is given by $n_1=0$ and a time independent phase $\phi_0(x)$ in $\phi(x,t)= \phi_0(x)+ \phi_1(x,t)$, found by solving  $\alpha \phi_{0,x}^2= 2\beta n_0$, that is $\phi_0(x)= \kappa x$ where $\kappa=  4 \sqrt{\pi a_\mathrm{scatt} n_0}$.
Recall that the strong interactions give $a_\mathrm{scatt}\to \infty$ which in turn makes $\kappa$ a large quantity.
The phase $\phi_0(x)$ is thus highly oscillating and $\phi_1$ can be considered as a slowly varying phase. 
The equation \eqref{n-eq} for the amplitude $n$ is then approximated at first order in $\kappa$, and together with the LDA, one obtains the wave equation  
\begin{align}
    n_{1,t} + 8\sqrt{\pi a_\mathrm{scatt} n_0} n_{1,x} = \hbar m^{-1} 4\pi a_\mathrm{scatt}  n_0\mathrm{Im}(p^*)\, .
    \label{a1-eq}
\end{align}
The HBE equation \eqref{HBE} together with the wave equation \eqref{a1-eq} forms the HRMB equation, the main object of this article. 


\section{The RMB equations}

In this section, we will analyse the integrability of the HRMB equation as a member of four integrable equations arising from the first negative flow of the AKNS hierarchy. 

\subsection{Four equations}

The system of equations \eqref{HBE} and \eqref{a1-eq} forms the HRMB equations. 
Using the following change of variables
\begin{align}
    \begin{split}
    Q&= \mathrm{Re}(p),\quad  P=  -\mathrm{Im}(p),\\
        N&=2\pi\frac{\hbar}{m}(2f+1)\quad\mathrm{and}\quad   E= n_0 + n_1\, , 
    \end{split}
\end{align}
the four RMB quations can be written together as 
\begin{align}
    \begin{split}
    cE_t + E_x &= \alpha P\\
    P_t &=  EN+\sigma_2 \omega_0 Q \\
    N_t &=  - \sigma_1 EP\\
    Q_t &= -\omega_0 P\, ,
    \end{split}
    \label{RMBs}
\end{align}
 where we have changed frames in \eqref{a1-eq} to absorb all constants and used the arbitrary speed $c$.
We introduced $\sigma_{1,2}= \pm 1$ that selects the two RMB for $\sigma_1=1$ and $\sigma_2= \pm 1 $ and the two HRMB equations for $\sigma_1 =-1$ and $\sigma_2=\pm 1$.  
They are all integrable, as shown below, but only the equation with $\sigma_1=1,\sigma_2=1$ has been derived before. 
Finding a physical interpretation for the RMB and the HRMB with $\sigma_2=-1$ remains an open problem.  

From \eqref{RMBs} we see that the generalised Bloch sphere is given by
\begin{align}
    P^2 + \sigma_2 Q^2 + \sigma_1 N^2= \eta\, ,
    \label{Bloch}
\end{align}
which is a hyperboloid when either or both of the $\sigma_{i}$ are negative.
The quantity 
\begin{align*}
    H = \frac{1}{2c} E^2 + \sigma_1 N\, , 
    \label{H}
\end{align*}
is also conserved by the RMB equation, provided the boundary conditions are periodic or vanishing.

Notice that setting $\omega_0= 0 $ recovers the Sine-Gordon equation from the RMB equations and the Sinh-Gordon equation from the HRMB equations. 
Indeed, for the HRMB equations, the change of variables $E= \phi_x$, $P= \sinh(\phi)$ and $N= \cosh(\phi)$ gives 
\begin{align}
    \phi_{xt}= 2\sinh(\phi)\, .  
\end{align}
This reduction is important, as after the KdV and NLS equations \cite{zakharov1972exact}, the Sine-Gordon equation was the next to be shown to be completely integrable 
\cite{caudreyPRL1973,ablowitz1973method,ablowitz1973nonlinear}.  

\subsection{Complete integrability}

We will now show that all of the RMB and HRMB equations are integrable by mean of the inverse scattering transform (IST).
For this we adopt a different convention for space and time variables, that is $t\leftrightarrow x$. 
We will also select the particular case of $c=0$ and $\alpha=1$ to simplify the exposition.
The spectral problem associated to these equations is the Zakharov-Shabat spectral problem \cite{zakharov1972exact, ablowitz1973nonlinear}\,; that is 
\begin{align}
    \begin{split}
    \Psi_x &= L_{\sigma_1}\Psi\\
    \Psi_t &= M_{\sigma_1,\sigma_2}\Psi\ ,,
    \end{split}
    \label{ZS}
\end{align}
where $\Psi= (\psi_1, \psi_2)^T$ is the scattering wavefunction and $L$ the spectral operator
\begin{align}
    L_{\sigma_1} = \lambda 
    \begin{bmatrix}
        i & 0\\
        0 & -i 
    \end{bmatrix} + 
    \begin{bmatrix}
        0 &E\\
        -\sigma_1 E & 0 
    \end{bmatrix}\, ,
    \label{L-matrix}
\end{align}
for the spectral parameter $\lambda$. 
Well-known equations such as the KdV or NLS equations can be written in such a spectral problem with the operator $M$ having only positive powers of $\lambda$. 
These would be the so-called positive AKNS hierarchy \cite{ablowitz1973nonlinear}. 
Here, we will use the negative part of the hierarchy, where the $M$ operator has negative powers of $\lambda$. 
It is given for the RMB ($\sigma_1=1$) and HRMB ($\sigma_1=-1,\sigma_2=1$) by  
\begin{align}
    M_{\sigma_1,+} = \frac{1}{2(\lambda^2-\omega_0^2)} \left( \lambda
    \begin{bmatrix}
        -iN & P\\
        \sigma_1 P & iN  
    \end{bmatrix} - 
    \omega_0
    \begin{bmatrix}
        0 & Q \\
        -\sigma_1 Q & 0
    \end{bmatrix}\right)\, ,
\end{align}
whereas the HRMB and RMB cases with $\sigma_2=-1$ have a different $M$ operator, given by
\begin{align}
    \begin{split}
    M_{\sigma_1,- } &= \frac{1}{2(\lambda^2-\omega_0^2)(\lambda-\omega_0)} \left ( i\lambda^2
    \begin{bmatrix}
        0 & P\\
        - \sigma_1 P & 0  
    \end{bmatrix}\right .-  \\ 
    &-\left .  
    \lambda \omega_0
    \begin{bmatrix}
        0 & Q \\
        \sigma_1 Q & 0
    \end{bmatrix}
    +i\omega_0^2
    \begin{bmatrix}
        N & 0\\
        0 & -N
    \end{bmatrix} \right )\, .
    \end{split}
\end{align}
The RMB and HRMB equations appear from computing the compatibility condition between the two equations in \eqref{ZS}, that is 
\begin{align}
    \partial_tL_{\sigma_1}- \partial_xM_{\sigma_1,\sigma_2} +[L_{\sigma_1},M_{\sigma_1,\sigma_2}]= 0\, .
    \label{ZCR}
\end{align}
This allows for the use of the IST by first solving the scattering problem, i.e., compute the eigenvalues of the scattering problem with the operator $L$, thus evolving them with the $M$ operator to finally reconstruct the solution by inverting the scattering problem. 
We will briefly use this method in the next section and just comment on the spectral problems here. 
In the case $\sigma_1=1$, the $L$ operator is anti-Hermitian, which means that the spectrum can have isolated eigenvalues in the case of vanishing boundary conditions, i.e. $E(\pm \infty)=0$.
In the hyperbolic case this operator is Hermitian, and so no discrete eigenvalues exist unless the boundary conditions are non-vanishing. 
This feature is also found in the nonlinear Schr\"odinger equation, where $\sigma=1$ corresponds to the focusing NLS, and $\sigma_1=-1$ to the defocusing case. 
The solitons in the latter equation are of a different type than the first and could be either dark or grey solitons, or even kinks, as in the HRMB equations -- see below.

An interesting feature of this spectral problem is that although the RMB and HRMB equations, with either $\sigma_2=\pm1$, share the same $L$ operator, the $M$ operator differs.
Because only the operator $L$ describes the shape of the solitons, the shapes are uniform on the sign of $\sigma_2$. 
The main difference between the two $M$ operators is in the position of the poles in the $\lambda$-plane. 
If $\sigma_2=1$, there are two simple poles $\lambda=\pm \omega$, and if $\sigma_2=-1$ there is a double pole at $\lambda= \omega_0$. 
Notice here that this is an arbitrary choice, and that $\lambda=-\omega_0$ could have also been the double pole. 
This is an unusual feature not present in the NLS equation which only contains positive powers of $\lambda$ in the $M$ operator. 
Shifting of the zeros by some parameter $\omega_0$ will only produce a gauge equivalent equation.

\subsection{Finite dimensional reduction}

The RMB equations \eqref{RMBs} have the same finite dimensional reduction for $c\neq 0 $, different from the original Bloch equation, or hyperbolic Bloch equation. 
The reduction is to assume that all the fields are constant in space, which consists of removing the spatial derivative in the equation for $E$, so it reduces to $cE_t= P$. 
In general, there is a constant of integration to take into account but we set it to $0$ for simplicity here.  
In \eqref{RMBs}, this equation is similar to the $Q$ equation, so we can set $E= -\frac{1}{c\omega_0} Q$ to obtain the finite dimensional system
\begin{align}
    \begin{split}
        P_t&= - \frac{1}{c\omega_0} QN + \sigma_2 \omega_0 Q\\
        N_t & = \frac{\sigma_1}{c\omega_0}QP\\
        Q_t &= - \omega_0 P\, . 
    \end{split}
    \label{ODE-reduction}
\end{align}
This system has the two independent conserved quantities
\begin{align}
    H&= \frac{\sigma_1}{2c\omega_0} Q^2 + \omega_0 N\label{H-quantity}\\
    C&= \frac12 P^2 + \frac{\sigma_1}{2}(N-\sigma_2c\omega_0^2)^2\label{C-quantity}\, . 
\end{align}
Notice that the Bloch sphere, or hyperboloid is not independent of $H$ and $C$. 

This system which we will call the ODE-RMB equation is exactly the non-dissipative Lorenz 63 model of \cite{lorenz1963deterministic}. 
This non-dissipative version of the Lorenz system is not well-studied in the literature, and we only refer to the monograph \cite{sparrow1982lorenz}, and in particular to the Appendix K, they treat the non-dissipative Lorenz system in the limit $\omega_0\to 0$, corresponding to the limit to the sin/sinh-Gordon equation. 
A similar reduction, but from the self-induced transparency equation to the complex Lorenz system, was derived and investigated by \cite{fowler1982complex,fowler1983real,gibbon1982real,gibbon1980derivation}. 

\begin{figure*}[htpb]
    \centering
    \subfigure[$\sigma_1=1,\sigma_2=1$: MI regime]{\includegraphics[scale=0.5]{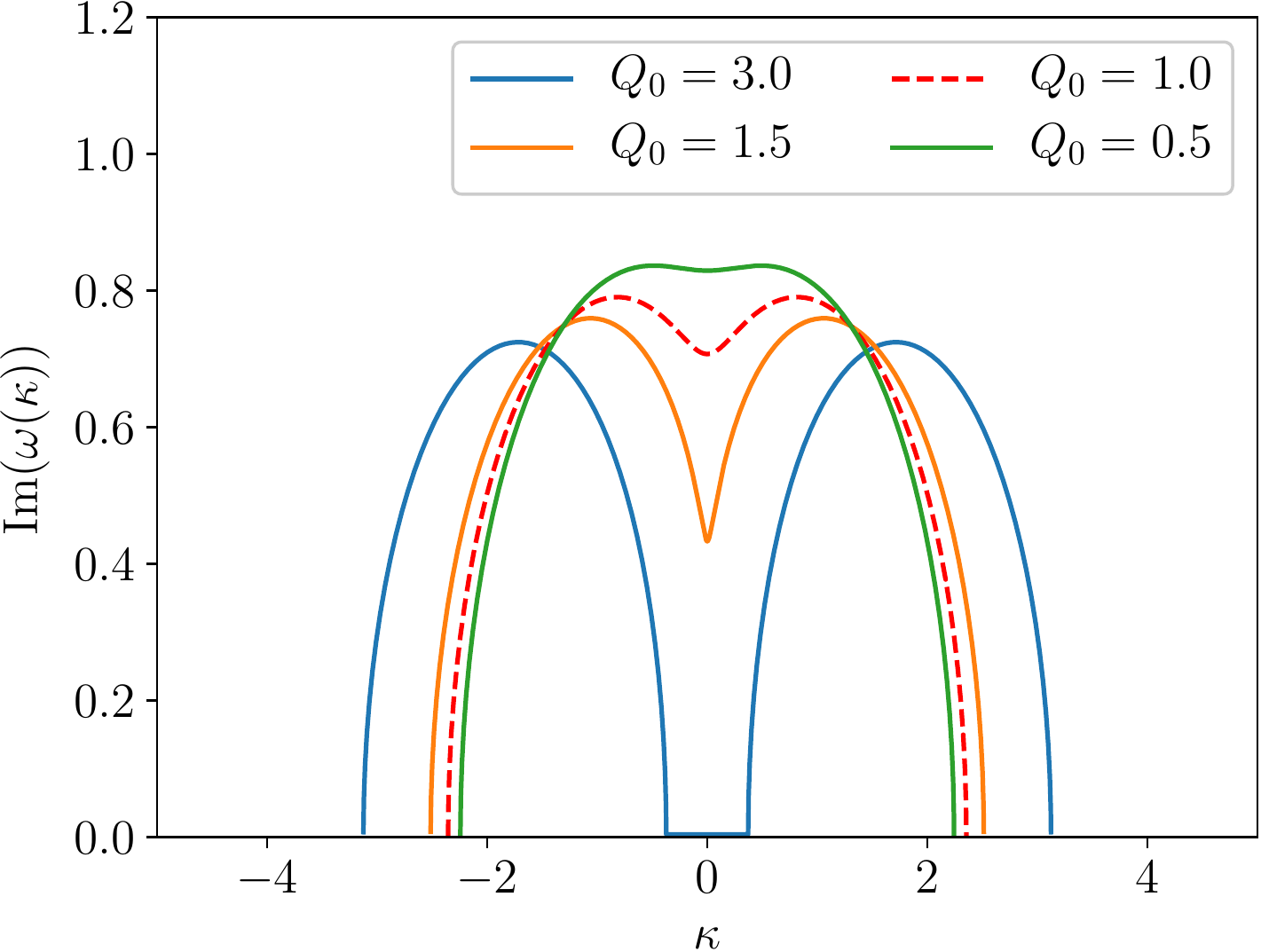}}
    \subfigure[$\sigma_1=1,\sigma_2=-1$: MI regime or unstable]{\includegraphics[scale=0.5]{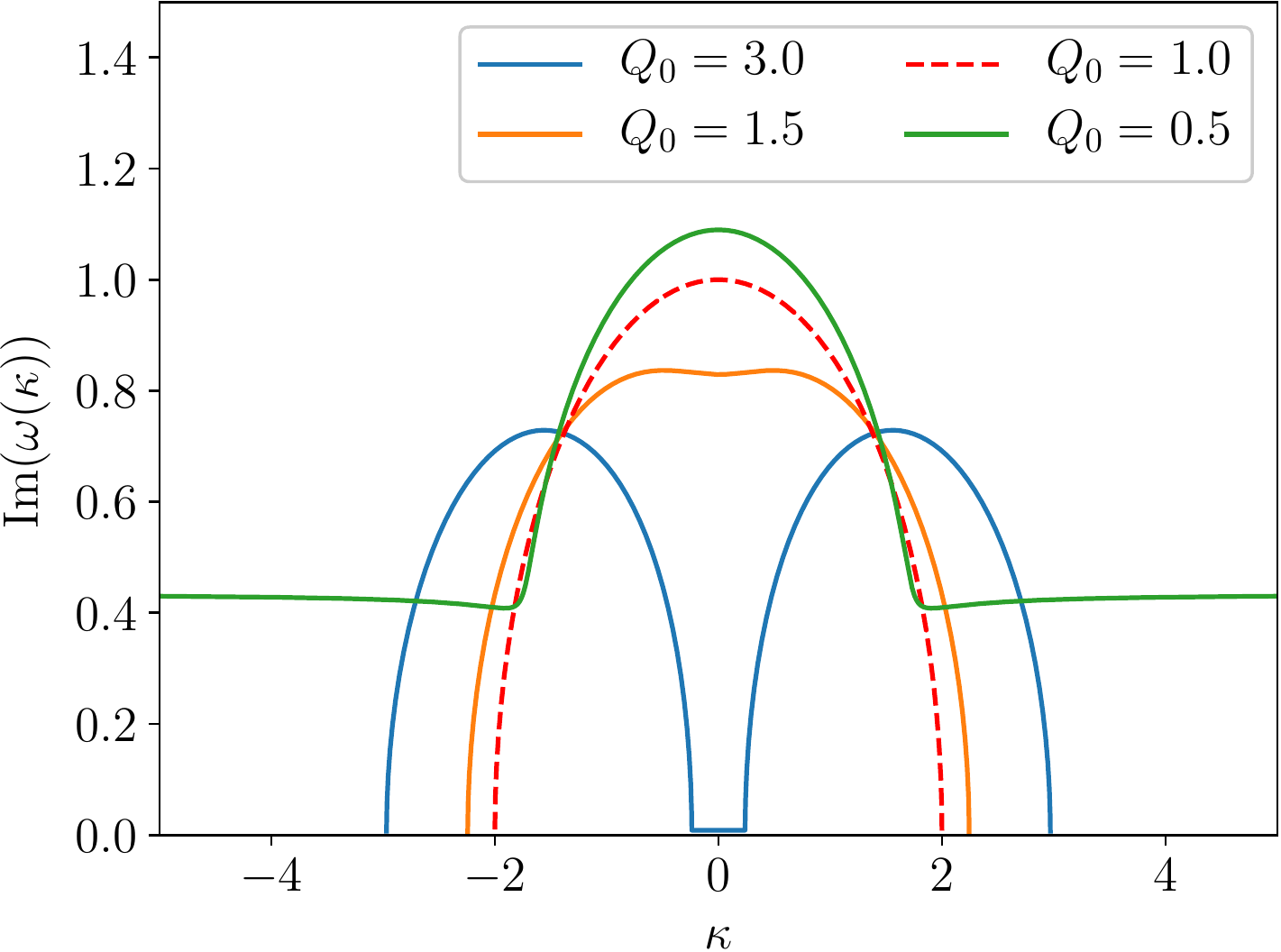}}
    \subfigure[$\sigma_1=-1,\sigma_2=1$: Stable or unstable]{\includegraphics[scale=0.5]{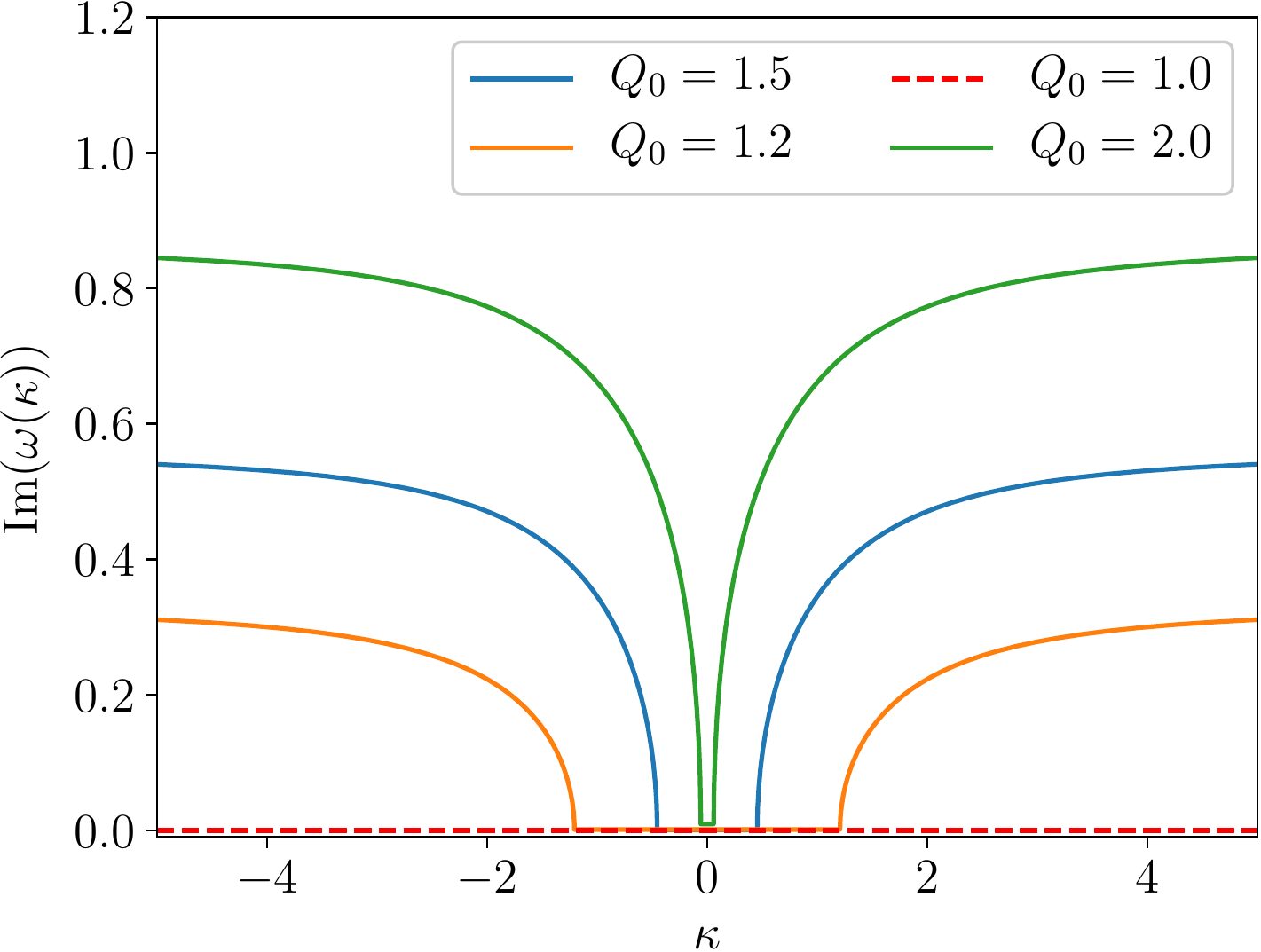}}
    \subfigure[$\sigma_1=-1,\sigma_2=-1$: Unstable]{\includegraphics[scale=0.5]{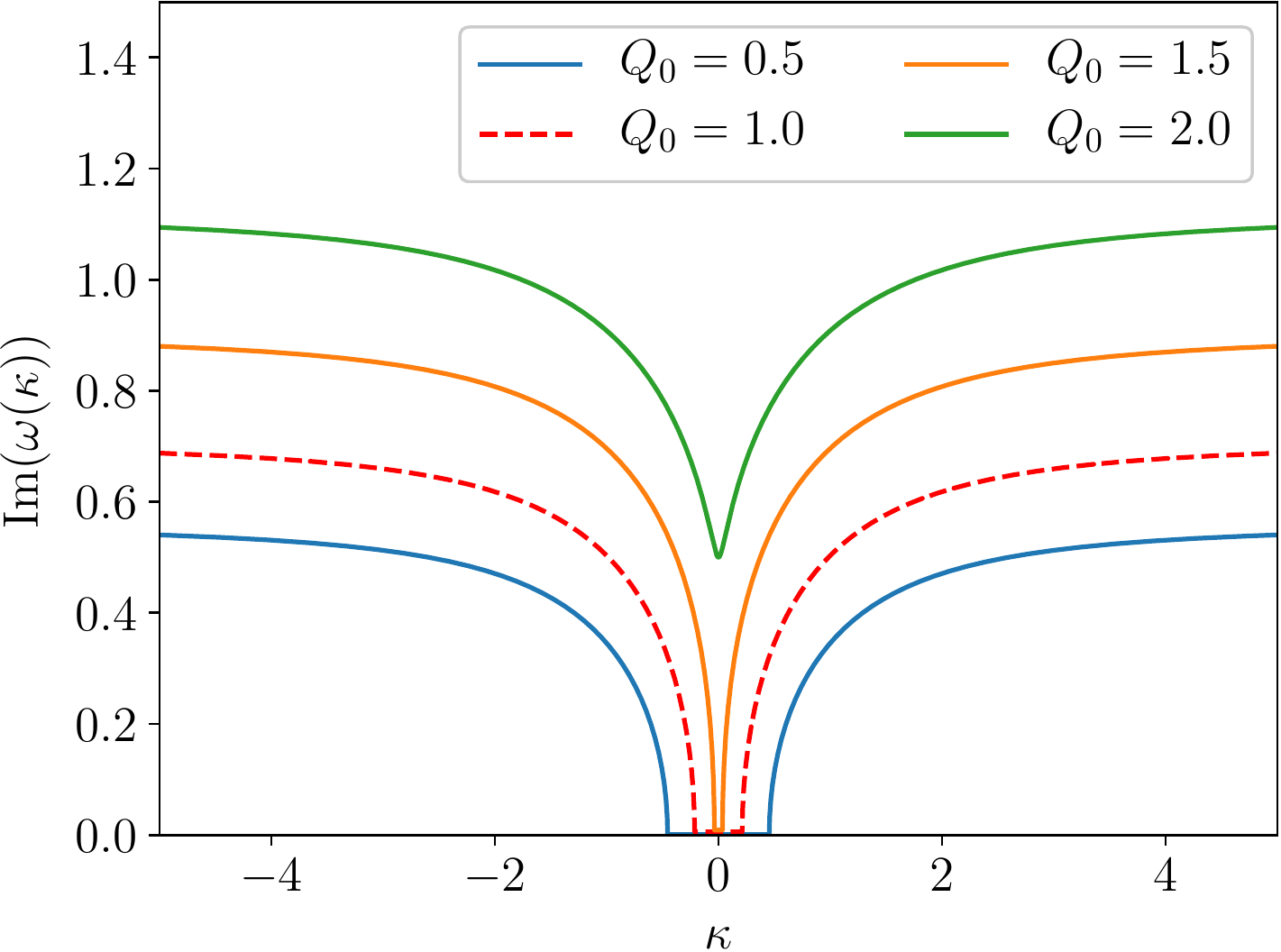}}
    \caption{ In this four panels, we display the instability gain of the four RMB equations, for some values of $Q_0$, $N_0= - \sigma_1$, $c=1$, $\omega=0.5$ and $\alpha=1$.
    We observe two various regimes which depends on the relative value of $Q_0$ and $N_0$ if $\sigma_1=-\sigma_2$.
     In particular for $\sigma_2=1$, there is a modulational instability regime for $\sigma_1=1$, and a stable regime for $\sigma_1=-1$. }
    \label{fig:MI}
\end{figure*}
\section{Solutions of the RMB equations}

We study here two classes of solutions of the four RMB equations, the constant solution related to the finite dimensional reduction \eqref{ODE-reduction} and one-soliton solutions.  

\subsection{Modulational instability}

The simplest solution of the RMB equations is the constant solution
\begin{align}
    \begin{split}
    Q(x,t)&= Q_0, \quad N(x,t)= N_0\\
    P(x,t)&= 0 \quad \mathrm{and}\quad E(x,t) = -\sigma_2\frac{Q_0}{N_0}\, , 
    \end{split}
    \label{stationary-sol}
\end{align}
for given constants $Q_0$ and $N_0$.
We study the linear stability of this family of solutions by linearising the RMB equation around this solution with perturbations of the form $N= N_0+\delta N$, for all four fields. 
We obtain
\begin{align}
    \begin{split}
        \begin{pmatrix}
            c\partial_t + \partial_x  & \alpha & 0 & 0 \\
            N_0 & \partial_t & -\sigma_2\frac{Q_0}{N_0} & \sigma_2 \omega_0\\
            0 & \sigma_1\sigma_2 \frac{Q_0}{N_0} & \partial_t & 0 \\
            0 & -\omega_0 & 0 & \partial_t 
        \end{pmatrix}
        \begin{pmatrix}
            \delta E\\
            \delta P\\
            \delta N\\
            \delta Q
        \end{pmatrix} = 0\, . 
    \end{split}
    \label{lin-RMBs}
\end{align}
We then assume that the perturbations are plane waves of the form $\delta N\propto \cos(\kappa x + \omega t)$ for all four fields. 
By a direct substitution into the linearised equations, we find that this ansatz solves the problem \eqref{lin-RMBs} if the following relation holds
\begin{align}
    \begin{split}
        -c  \omega^3 - \kappa \omega^2 + &\left (c\omega_0\frac{\sigma_2N_0^2 + \sigma_1Q_0^2}{ N_0^2} - \alpha N_0 \right ) \omega \\
        &+ \kappa \omega_0\frac{\sigma_2 N_0^2 + \sigma_1 Q_0^2}{N_0^2}= 0\, .
    \end{split}
\end{align}
We will solve this equation numerically to obtain the dispersion relation $\omega(\kappa)$ for all values of the parameters. 
One can already see that for $\sigma_2 = -\sigma_1$, the choice $N_0^2=Q_0^2$ will correspond to a transition between two different regimes. 

We plot in Figure~\ref{fig:MI} the instability gain of the four RMB equations varying the value of $Q_0$, for $N_0= -\sigma_1$, $c=-1$, $\omega_0=0.5$ and $\alpha=1$. 
When $\sigma_1 = - \sigma_2$, we observe a transition at $Q_0=N_0$ from a stable to an unstable regime. 
For $\sigma_1=1$, the stable regime has a band of modulational instability for low wavenumbers, and for $\sigma_1=-1$ the solution is stable for all wavenumbers. 
In more details, we have the following analysis of the Figure~\ref{fig:MI}.
\begin{enumerate}
    \item[(a)] $\sigma_1= 1,\sigma_2=1$: Instabilities for low wavenumbers, corresponding to a modulational instability (MI). 
    Depending on the value of the parameters, and in this case for large $Q_0$, a stable low-frequency region can exist. 
    Increasing $c$ would have a similar effect.  

    \item[(b)] $\sigma_1=1,\sigma_2=-1$: There are two cases:
        \begin{itemize}
            \item if $Q_0^2\leq N_0^2$, MI regime, similar to the previous case $\sigma_1=\sigma_2=1$;  
            \item if $Q_0^2> N_0^2$, MI regime together with a smaller amplitude instability for all frequencies.
        \end{itemize}

    \item[(c)] $\sigma_1=-1,\sigma_2=1$: There are two cases:
        \begin{itemize}
            \item if $Q_0^2\leq N_0^2$, the constant solution is stable for all frequencies;
            \item if $Q_0^2> N_0^2$, the constant solution is unstable for all frequencies, except for a small region of low frequencies. 
        \end{itemize}

    \item[(d)] $\sigma_1=-1,\sigma_2=-1$: the solution is unstable for almost all frequencies and for any values of $Q_0$. 
        As in the previous case, there is a small stable band at low frequencies, for small values of $Q_0$. 
\end{enumerate}

\begin{figure}[htpb]
    \centering
    \subfigure[Bloch sphere with $\sigma_1=1,\sigma_2=1$]{\includegraphics[scale=0.43]{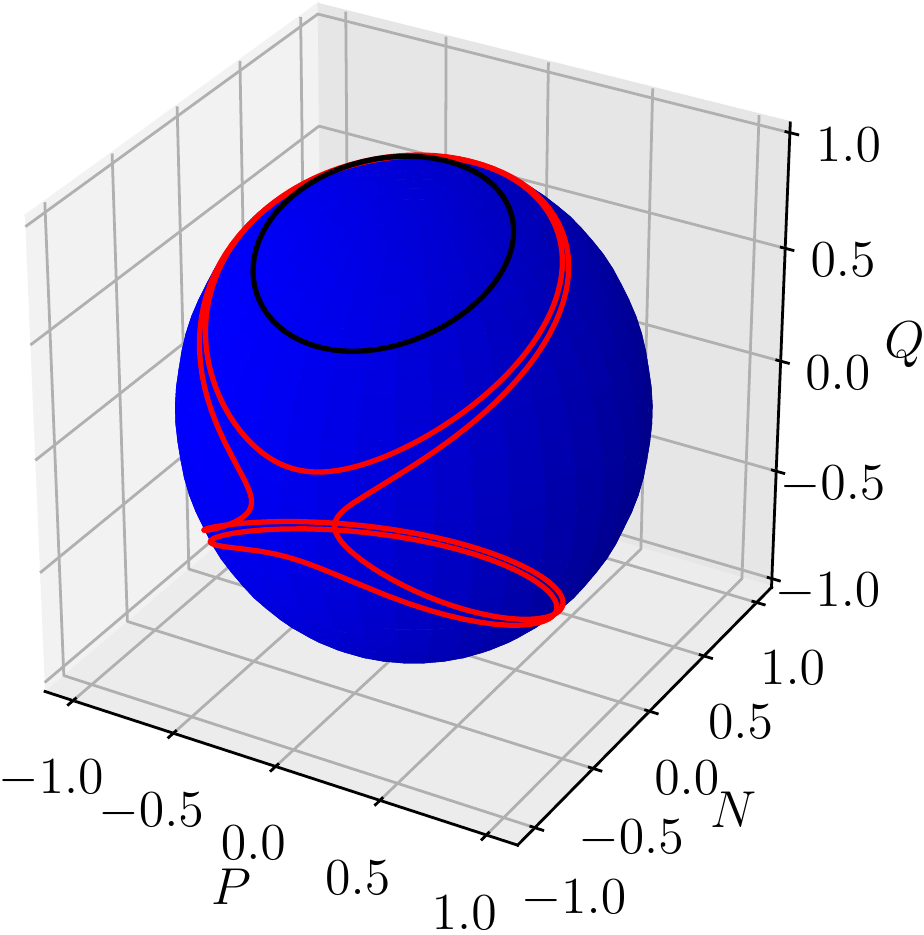} \label{fig:L63p}}
    \subfigure[Bloch hyperboloid with $\sigma_1=-1,\sigma_2=-1$]{\includegraphics[scale=0.43]{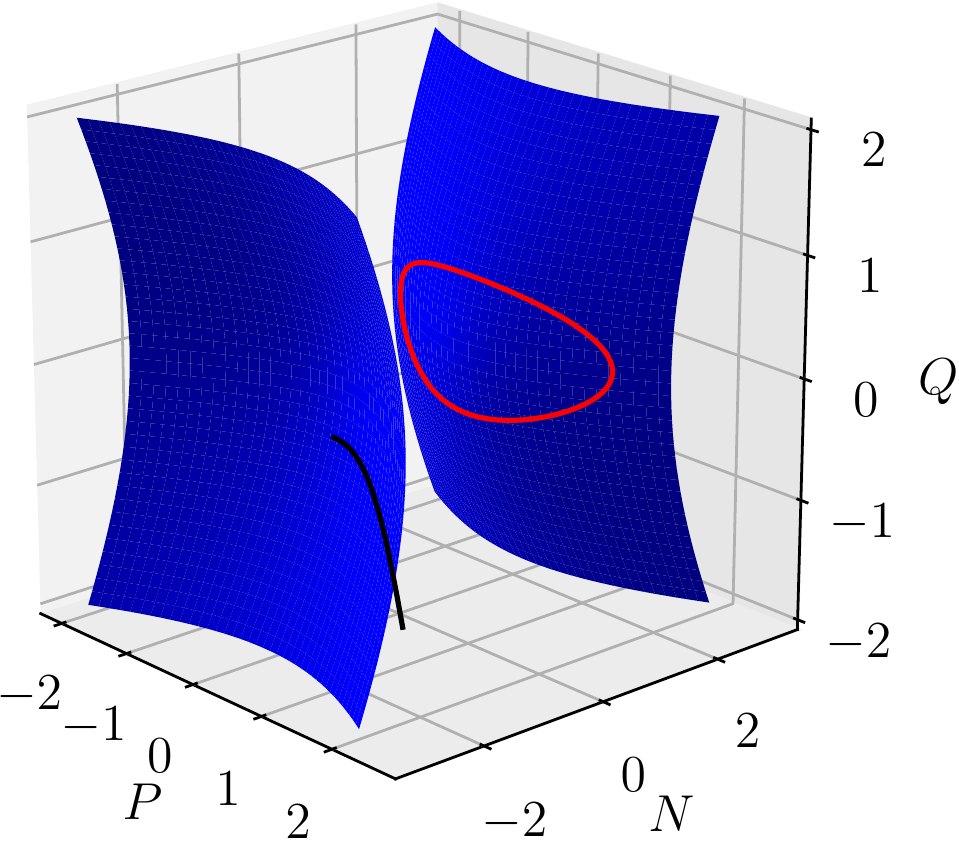}\label{fig:L63h}}
	\caption{In this figure, we display typical orbits of ODE-RMB equation \eqref{ODE-reduction} on the Bloch sphere (left) and Bloch hyperboloid (right), obtained as a finite dimensional reduction of the RMB equations. 
    On the left panel, the standard RMB equation with $\sigma_1=\sigma_2=1$ of quantum optics has solutions restricted to the Bloch sphere, with stable equilibrium in the centre of the black orbit, and unstable saddle point near the centre of the red orbit.
    This position corresponds to the MI regime of the RMB equation. 
    On the right panel, we show two orbits of the reduction of the HRMB equation of strongly correlated Bose-Einstein condensates, which are, in this case, restricted to a two sheet hyperboloid. 
    The stationary solution at the centre of the red orbit is stable, but unstable near the black orbit. 
    These two scenarios correspond to the stability and instability of the full HRMB equation. }
\end{figure}

The finite dimensional reduction to the ODE-RMB equation \eqref{ODE-reduction} can be used to understand these regimes of instability.
First, in the case $\sigma_1= \sigma_2=1$, the dynamics takes place on the Bloch sphere \eqref{Bloch} intersected by a cylinder in the $Q$ directions (given by $C$ in \eqref{C-quantity}), or parabolic sheet in the $P$ direction (given by $H$ in \eqref{H-quantity}). 
The physical stationary point corresponds to $Q=0$, as we did not include a constant of integration in the finite dimensional reduction. 
This fix point is hyperbolic, thus unstable and corresponds to the MI regime. 
The other fix points are $P=0,\quad  Q= \pm Q_0\quad \mathrm{and}\quad N= \sigma_2 c \omega_0^2$, which are stable but not physical. 
We show some stable and unstable orbits in Figure~\ref{fig:L63p}.

In the case when $\sigma_1=-1,\sigma_2=1$, the dynamics takes place at the intersection between a hyperboloid in the $N$ direction given by \eqref{Bloch} and a parabolic sheet in the same direction (given by $H$ in \eqref{H-quantity}).
The hyperboloid can have one or two sheets, depending on the relative value of $Q$ and $N$. 
In the two sheets case ($Q_0^2\leq N_0^2$), selecting $N_0=1$ restricts the dynamics to the stable sheet, where the parabola is bounded from below. 
If the hyperboloid is of one sheet, or $N_0=-1$, the trajectory can reach the unbounded region of the parabola. 
This second case corresponds to an unstable regime of the HRMB equation. 
In Figure~\ref{fig:L63h}, we show a stable and unstable orbit of the ODE-RMB equation \eqref{ODE-reduction}.  

For the HRMB, with $\sigma_1=-1$ and $\sigma_2=1$, the inequality $Q_0^2\leq N_0^2$ corresponds to a condition that the transition rate between the BECs atoms and the excited atoms should be small enough compared to the number of excited atoms. 
In physical variables, we have the condition that
\begin{align}
    |\mathrm{Re}(p)| < 2\pi \frac{\hbar}{m}|2f +1|\,. 
\end{align}
For the RMB equation, this regime of MI is to be expected, as this equation is similar to the focussing NLS, which is the most important example of modulational instability. 
In the nonlinear regime of the modulational instability, when the linear approximation of the integrable equation is not valid anymore, the nonlinearities prevent the solution to blow up and form a train of interacting pulses. 
This nonlinear solution can be understood as a regime of integrable turbulence and can produce high amplitude waves, called rogue waves. 
We only refer to \cite{zakharov2009turbulence} for the notion of integrable turbulence and to \cite{agafontsev2015integrable} for the formation of rogue waves for the NLS equation. 

\subsection{Bright solitons}

\begin{figure}[htpb]
    \centering
    \subfigure[$E$ field for $\sigma_2=1$]{\includegraphics[scale=0.45]{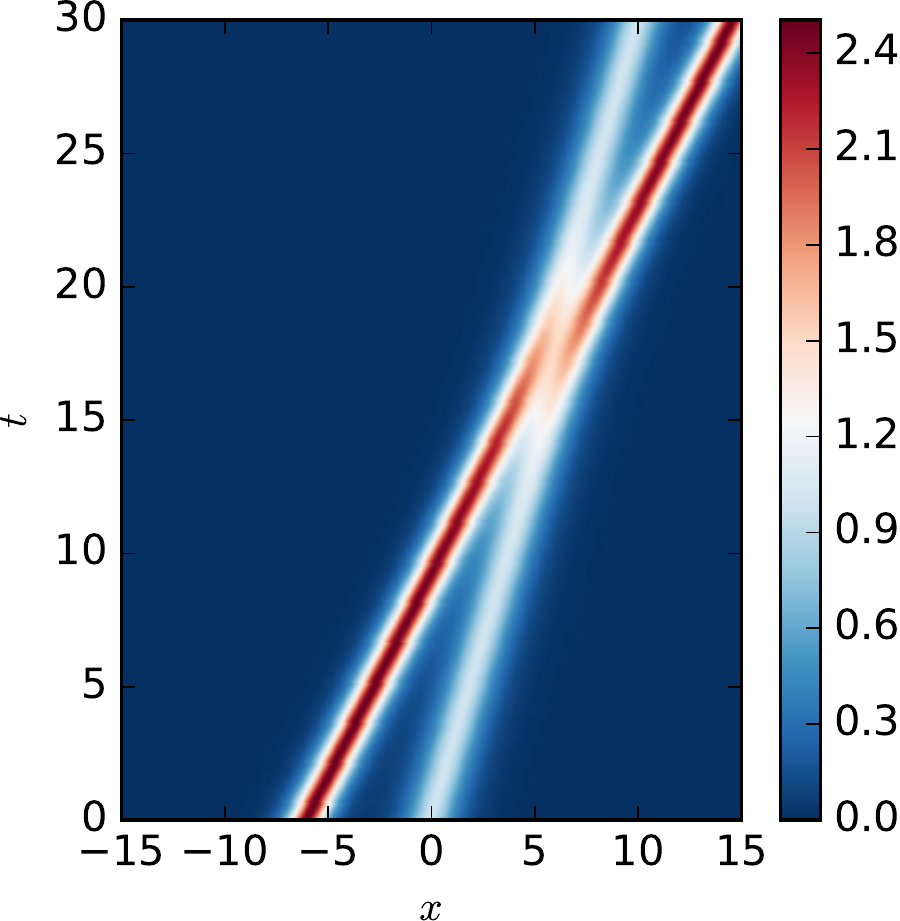}}
    \subfigure[$E$ field for $\sigma_2=-1$]{\includegraphics[scale=0.45]{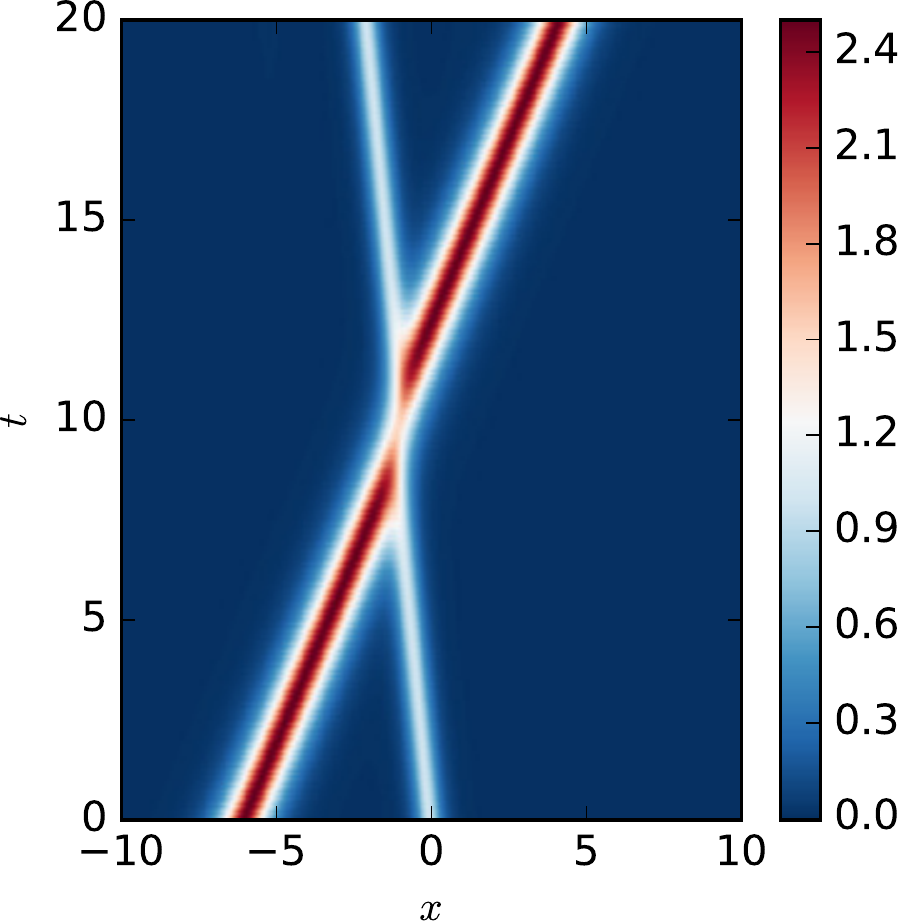}}
    \caption{We display the $E$ field of two collisions of the RMB solitons for $\sigma_1=1$ and $\sigma_2=1$ on the left panel and $\sigma_2=-1$ on the right panel. 
    The initial conditions have the same parameters $A=1, 2.5$ and $\omega_0=0.6$ in \eqref{sech-solution}, taken such that the overtaking collision corresponds to a head-on collision for $\sigma_2=-1$. 
	Both noticeably display a different phase shift after the collision, which can be estimated from the theory of integrable systems.}
    \label{fig:RMB-soliton}
\end{figure}

As already mentioned, soliton solutions can be derived with the IST method, but here we will find them by simply using the travelling wave ansatz $E(x,t)= E(t-c^{-1}x)$ for a constant parameter $c$. 
We find the following ODEs when using the boundary conditions $N(\pm\infty)= N_\infty$ and $P(\pm \infty)= Q(\pm\infty) = E(\pm \infty) = 0$ 
\begin{align}
    E_{xx}= -E\left ( \frac12 \sigma_1 E^2 +  cN_\infty  + \sigma_2\omega_0^2\right )\, .
    \label{Exx}
\end{align}
The sign in front of the $E^3$ terms changes the type of solution, from $\mathrm{sech}$-profile to a $\mathrm{tanh}$-profile. 
For the solution of the RMB equations, we obtain with $N_\infty= - 1$, the kink solution
\begin{align}
    E(x,t)= E_0\, \mathrm{sech}\left (\frac12 E_0 \left (t- \frac{4}{ E_0^2 + 4\sigma_2\omega_0^2}x\right )\right )\, .
    \label{sech-solution}
\end{align}

We display the soliton \eqref{sech-solution} of the RMB equations with $\sigma_1=1$ and $\sigma_2=\pm 1$ in Figure~\ref{fig:RMB-soliton} where we numerically computed two collisions with the same initial conditions for both cases.
We have used the Python package Dedalus \cite{burns2016dedalus} to perform these simulations.
We used $E_0=1$ and $E_0=2.5$ with $\omega_0=0.6$ in \eqref{sech-solution}. 
Notice that the initial conditions are different as they depend on the value of $\sigma_2$. 
The standard RMB equation with $\sigma_2=1$ has only right going solitons overtaking each other. 
For $\sigma_2=-1$, solitons have opposite directions and we observe a head-on collision, scenario which does not appear in the standard RMB equation. 
These simulations were obtained by directly solving the RMB equation with $E_x+ E_t= P$ for the $E$ equation in \eqref{RMBs}.

\subsection{Kink solitons}

A direct integration of the travelling wave ODE \eqref{Exx} would give the kink soliton of the form
\begin{align}
    E(x,t) = \pm E_\infty\, \mathrm{tanh}\left ( 2E_\infty \left ( t- \frac{N_\infty  }{E_\infty^{-2}/2-\sigma_2\omega_0^2}x \right ) \right )\, , 
	\label{tanh-sol}
\end{align}
but this is not a valid solution of the HRMB equation which requires particular boundary conditions in order to be constant at $\pm \infty$. 
In particular, we need $Q(\pm \infty) = \frac{E(\pm \infty) N_0}{\sigma_2 \omega_0}$, which takes opposite values at $\pm \infty$. 
Using this boundary condition, we obtain the travelling wave ODE
\begin{align}
	E_{xx}= E\left ( \frac12E^2 -  cN_\infty  - \sigma_2\omega_0^2\right ) + \mathrm{sgn}(x) c E_\infty N_0\, ,
	\label{Exx-HRMB}
\end{align}
where the last constant is positive for $x>0$, and negative for $x<0$.  
Finding explicit solutions of this equation is out of the scope of this work, but one can see that they remain close to the tanh solution \eqref{tanh-sol}.

\begin{figure}[htpb]
    \centering
	\subfigure{\includegraphics[scale=0.55]{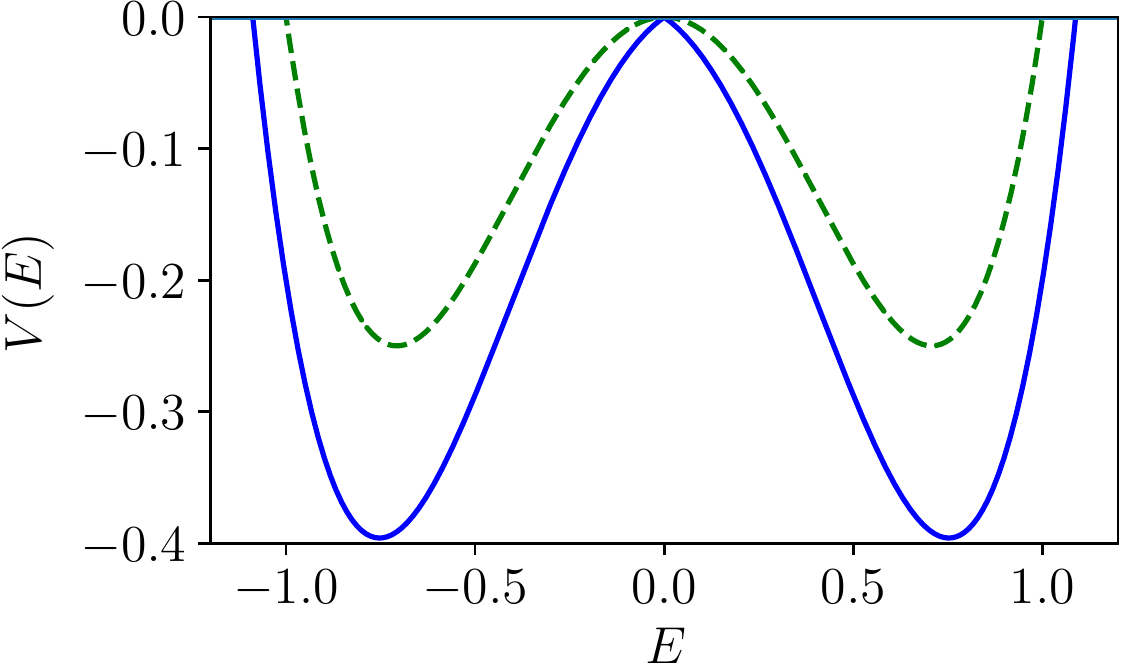}}
	\subfigure{\includegraphics[scale=0.55]{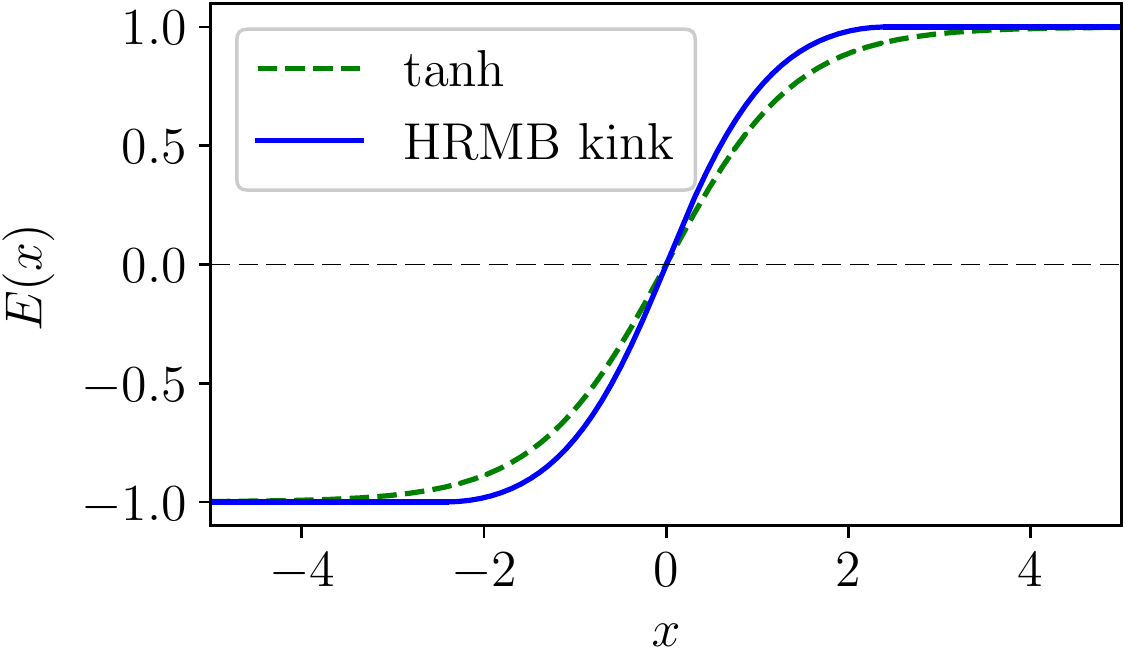}}
	\caption{In the top panel we display the potential function $V(E)$ in \eqref{VE}  describing the travelling wave solutions of the HRMB equation in blue, and compare with the potential corresponding to the tanh solution in dashed green. 
	In the bottom panel, we compare the solutions, numerically integrated from \eqref{newton}.}  
    \label{fig:kink-soliton}
\end{figure}
In Figure~\ref{fig:kink-soliton}, we compare the tanh solution with a numerical solution of \eqref{Exx-HRMB}, obtained by considering the following simplified model for \eqref{Exx-HRMB}
\begin{align}
	E''(x) &= \nabla V(E(x))\, ,
	\label{newton}
\end{align}
with the potential function 
\begin{align}
	V(E(x)) &= E(x)^4- E(x)^2- 0.2 \mathrm{sgn}(x) E(x)\, . 
	\label{VE}
\end{align}
Notice that without $\mathrm{sgn}(x)$, the linear additional term makes this potential non-symmetric with respect to the origin. 

\section{Conclusion}

In this work, we have derived the hyperbolic RMB equation from the theory of strongly correlated Bose-Einstein condensate to model the interaction of the BEC with its evaporated atoms. 
The HRMB equation have been obtained as the result of several approximations of a complete physical model and has the remarkable property of being completely integrable via the inverse scattering transform. 
This equation is in fact a member of the negative flow of the AKNS hierarchy together with three other equations, one being the original RMB equation of quantum optics. 
After showing the integrability of these four equations, we have studied some of their solutions including the stability of constant solutions and the one-soliton solutions. 
The HRMB equations turn out to have a stable flat background only if the rate of interaction between the BEC and the normal component is small enough compared to the number of evaporated atoms. 

This work also raises several open questions, left for future work. 
The first includes further studies of this equation in the context of integrable systems, such as the derivation of more solutions, as well as a study of their properties, including an understanding of the fact that the Lax pair contains poles in the complex plane of the spectral parameter. 
The explicit form of the kink soliton and the multi-kink solitons would also be an interesting challenge using the theory of IST. 
The second open problem is more physical and will require more work to assess the validity of the approximations that have been made\,: for example, can they be shown to be consistent with certain experimental conditions so that the solitons could be observed for a long enough period of time?
Other research directions include the application of the IST method for other solutions, a complete study of the modulational instability regime, with the possible existence of rogue waves, the connection with the Lorenz 63 model and the possible physical application of the RMB equations with $\sigma_2=-1$. 


\begin{acknowledgments}
    We acknowledge, with thanks, discussions with R. Barnett, M. Kira, D. Holm, A. Hone and A. Newell. 
	The first author acknowledges partial support from an Imperial College London Roth Award and from the European Research Council Advanced Grant 267382 FCCA.
\end{acknowledgments}

\bibstyle{apsrev4-1.bst} 
\bibliography{HRMBbiblio}

\end{document}